\documentclass[numberedappendix,apj]{emulateapj}
\usepackage{amssymb,amsmath}
\usepackage{natbib}
\usepackage{graphicx}
\shorttitle{Velocity anisotropy}
\shortauthors{Esquivel \& Lazarian}
\slugcomment{Draft version, \today}

\defcitealias{2002ApJ...564..291C}{CLV02}
\defcitealias{1995ApJ...438..763G}{GS95}
\defcitealias{2002ASPC..276..182L}{LPE02}
\defcitealias{2005ApJ...631..320E}{EL05}

\begin{document}

\title{Velocity Anisotropy as a Diagnostic of the Magnetization \\
 of the Interstellar Medium and Molecular clouds} 

\author{
  A. Esquivel\altaffilmark{1} \& A. Lazarian\altaffilmark{2}
}

\affil{
\altaffilmark{1}Instituto de Ciencias Nucleares, Universidad Nacional
Aut\'{o}noma de M\'{e}xico,\\ Apartado Postal 70-543, 04510 M\'{e}xico
D.F., M\'{e}xico\\ 
\altaffilmark{2}Astronomy Department, University of Wisconsin-Madison,
475 N. Charter Street, Madison, WI 53706-1582, USA
}

\email{esquivel@nucleares.unam.mx, lazarian@astro.wisc.edu}

\begin{abstract}
We use a set of magnetohydrodynamics (MHD) simulations of
fully-developed (driven) turbulence to study the anisotropy in the
velocity field that is induced by the presence of the magnetic
field. In our models we study turbulence characterized by sonic Mach
numbers $M_s$ from $0.7$ to $7.5$, and  Alfv\'en Mach numbers from
$0.4$ to $7.7$. These are used to produce synthetic observations
(centroid maps) that are analyzed. To study 
the effect of large scale density fluctuations and of white noise we
have modified the density fields and obtained new centroid maps, which
are analyzed.  We show that restricting the range of scales at which
the anisotropy is measured makes the method robust against such
fluctuations. We show that the anisotropy in the structure function of
the maps reveals the direction of the magnetic field for $M_A\lesssim
1.5$, regardless of the sonic Mach number.  We found that the degree
of anisotropy can be used to determine the degree of magnetization
(i.e. $M_A$) for $M_A\lesssim 1.5$. To do this, one needs an
additional measure of the sonic Mach number and an estimate of the LOS
magnetic field, both feasible by other techniques, offering a new
opportunity to study the magnetization state of the interstellar
medium. 
\end{abstract}


\section{Introduction}
\label{sec:introduction}

It is well accepted that turbulence plays a central role in the
dynamics and transport phenomena in the interstellar medium
(ISM). For an overview we refer the reader to the recent reviews by
\citet{2004ARA&A..42..211E, 2004RvMP...76..125M, 2007prpl.conf...63B,
  2007ARA&A..45..565M}, and references therein. 
Moreover, turbulence in the ISM is magnetized, so in order to
understand its properties one has to address those of the magnetic
field as well. 

The presence of a magnetic field introduces a preferential direction
of motion for the charged particles and in consequence makes the
turbulent cascade anisotropic, which has been known for some time now
\citep{1982PhyS....2...83M, 1984ApJ...285..109H}. In a turbulent
magnetized medium the kinetic energy of large-scale motions is larger
than those at small-scales, but the local magnetic field  (thus
the magnetic energy) is comparable at all scales. Therefore at
large-scales the magnetic field does not dominate dynamically, but it 
becomes more important as we go into smaller scales. The result are
elongated eddies, which become more elongated as the energy cascade
goes down to smaller scales \citep*[][henceforth
\citetalias{2002ApJ...564..291C}]{2000ApJ...539..273C,
  2001ApJ...554.1175M,  2002ApJ...564..291C}.  
Recent decades have seen important progress in the understanding of the
magnetic turbulence. A pioneering study is the self-consistent
model of magnetohydrodynamic (MHD) turbulence of \citet[][hereafter
\citetalias{1995ApJ...438..763G}]{1995ApJ...438..763G} where the
concept of a scale dependent anisotropy is built in\footnote{The initial
  publication did not have the concept of the local magnetic field
  built in. In fact, the closure relations in the paper were written
  in the mean field frame of reference. The correct understanding that
  the eddies can be described only in the system of reference related
  to the local magnetic field of the eddies in question was obtained
  later (\citealt{1999ApJ...517..700L,2000ApJ...539..273C,
    2001ApJ...554.1175M}; \citetalias{2002ApJ...564..291C})}.
The \citetalias{1995ApJ...438..763G} model was later supported by the
numerical simulations of \citet{2000ApJ...539..273C,
  2001ApJ...554.1175M}; \citetalias{2002ApJ...564..291C} . The ideas on 
compressible turbulence sketched in \citetalias{1995ApJ...438..763G}
have proved profound, and have been supported and developed in further
studies \citep{2001ApJ...562..279L, 2003MNRAS.345..325C,
  2010ApJ...720..742K}.
While some aspects of \citetalias{1995ApJ...438..763G} scaling were
claimed to be controversial and not consistent with  numerical simulations,
e.g. the predicted Kolmogorov-type index of the spectrum, later
studies \citep[e.g.][]{2010ApJ...722L.110B} have revealed the
limitations of the numerical simulations which induced these
controversies. All in all, we believe that the
\citetalias{1995ApJ...438..763G} model (with the adjustments and
improvements introduced in subsequent publications) provides the best
representation of the MHD turbulence statistics. 

The \citetalias{1995ApJ...438..763G} model is well described in many
reviews \citep[see][]{2005PhST..116...32L} and we do not dwell upon its
details. We just mention that the qualitative understanding of the
model can be obtained if one assumes that mixing motions perpendicular
to the local magnetic field create a turbulent Kolmogorov-like
cascade. The eddies are elongated along the local direction of
magnetic field and the relation between the parallel and perpendicular
scales of the eddies is given by the so-called ``critically balance,''
which is reflected in the equality of the eddy turnover time and the
timescale of propagation of Alfv\'en waves along the magnetic field of
the eddy. The tensor describing the turbulent magnetic field is presented in
\citetalias{2002ApJ...564..291C}.

The underlying anisotropies of turbulence result in the anisotropies
of the observed statistics of turbulence. However, the anisotropy of
interstellar turbulence which is accessible to the observer averaging
emission along the line of sight crossing a turbulent volume is
different from the \citetalias{1995ApJ...438..763G} predictions. 
In the ``global'' system of reference, related to the mean magnetic
field, the anisotropy is no longer-scale dependent but is determined
by the anisotropy of the largest eddies. A suggestion of using such
anisotropy to study turbulence and the direction of the mean magnetic
field was made \citet*[][henceforth
\citetalias{2002ASPC..276..182L}]{2002ASPC..276..182L} , where the 
feasibility of such studies was illustrated with synthetic spectral
line emission maps obtained via MHD turbulence simulations. The
anisotropy is readily evident from two-point statistics 
(e.g. correlation/structure functions and power-spectra), where more
power (larger dispersion) is concentrated in the direction
perpendicular to the magnetic field \citep[see for
instance][]{2003MNRAS.342..325E, 2003ApJ...590..858V}. 
The aforementioned studies provided the framework for the observational
studies in \citet{2008ApJ...680..420H}.

In this paper we revisit the velocity anisotropy from numerical
simulations and synthetic observations. In particular, we study how
such anisotropy depends on the global Alfv\'enic \emph{and} sonic
Mach numbers of the turbulence.

If one knows the Alfv\'en Mach number, then by knowing the turbulent
velocity dispersion, an estimation of the media magnetization (i.e. its 
Alfv\'en speed) is possible. The determination of the Alf\'en speed, which 
is a crucial parameter for theoretical and numerical models, has
been a great challenge for current techniques that probe the ISM.

We describe our MHD simulations in \S 2, and the
way of measuring velocity anisotropy in observations in \S 3. Results
of our study, namely the degree of anisotropy obtained as a
function of Alfv\'enic and sonic Mach numbers are presented in \S
4. We finish with a discussion and provide a summary in \S 5.

\section{MHD Models}
\label{sec:model}

We use a set of three-dimensional MHD simulations of fully-developed
(driven) turbulence to produce maps of velocity centroids and of
the average LOS velocity\footnote{The average LOS velocity traces the
  statistics of velocity in the same manner that column density traces
  density. However, it can not be obtained directly from observations,
  while column density can. In previous works we have termed it
  ``integrated velocity'' \citep[e.g. ][]{2003ApJ...592L..37L,
    2005ApJ...631..320E}.}.

The simulations presented in this work were obtained solving the ideal
MHD equations in a periodic box,
\begin{eqnarray}
\frac{\partial \rho}{\partial t}+\mathbf{\nabla \cdot}
\left(\rho\mathbf{v} \right) = 0, \label{eq:continuity}\\
\frac{\partial \rho\mathbf{v}}{\partial t}+\mathbf{\nabla
  \cdot}\left[\rho\mathbf{vv}+\left(p+\frac{B^2}{8\pi}\right)\mathbf{I}
  -\frac{1}{4\pi}\mathbf{BB} \right]=\mathbf{f},\label{eq:momentum}\\
\frac{\partial \mathbf{B}}{\partial t}-\mathbb{\nabla
  \times}\left(\mathbf{v \times B} \right)=\mathbf{0},\label{eq:induction}
\end{eqnarray}
with an isothermal equation of state ($p=c_s^2\rho$, where $p$ is the
gas pressure, $c_s$ the sound speed, and $\rho$ the mass density) and
the additional constraint of $\mathbf{\nabla\cdot B}=0$, achieved with
a constrained transport (CT) algorithm (see e.g. \citealt{2000JCoPh.161..605T}).
The integration method is a second-order-accurate hybrid essentially
nonoscillatory (ENO) scheme \citep[see][]{2002PhRvL..88x5001C}. In
order to avoid spurious osscilations, the code switches from a ENO
weighted scheme \citep[][where variables are
smooth]{1999JCoPh.150..561J} to a convex ENO scheme \citep[][where
strong gradients are found]{1989JCoPh..83...32S}. The time is marched
with a two stage Runge-Kutta method. The source 
term at the right hand side of equation (\ref{eq:momentum}) is a
large-scale pseudo-random driving. Such driving  is purely solenoidal
and is 
performed in Fourier space at a fixed wave number $k=2.5$ (a scale of
$1/2.5$ of the computational domain). For details about the driving see
\citet{2003MNRAS.345..325C} and \citet{2007ApJ...658..423K}.
Initially $\rho=1$ and the Alfv\'en speed $v_A=\vert\mathbf{B}\vert
/\sqrt{4\pi \rho}$ depends on the model. 
In stationary state the rms velocity is also close to unity
($v_{\mathrm{rms}}\sim 0.7$). 
The parameters that define the experiments used are the sonic and
Alfv\'enic Mach numbers at the injection scale
$M_s\equiv \left\langle V_L/c_s \right\rangle$ and 
$M_A\equiv \left\langle V_L/v_A \right\rangle$, respectively,  where
$V_L=v_{\mathrm{rms}}$ is the turbulent velocity at the injection scale, 
and $\langle \ldots \rangle$ stands for average
over the entire computational domain. These parameters can be
controlled by the values of the gas pressure and the Alvf\'en speed at
the beginning of the simulation, $\langle P_{gas,0}\rangle$ and
$v_{A,0}$, respectively. In Table  \ref{tab:models} we list the
numerical experiments that cover sub-sonic and super-sonic regimes,
combined with different intensities of magnetic field 
(yielding sub-Alfv\'enic and super-Alfv\'enic regimes as well).
We have to mention that the value of $v_{rms}$ in stationary state
varies from model to model, the Mach numbers presented in the table
are measured from the output of the simulations. 

\begin{deluxetable}{lccccc}
\tablecaption{Parameters of the MHD simulations. \label{tab:models}}
\tablewidth{0pt}
\tablehead{
    \colhead{Model}
  & \colhead{$v_{A,0}$}
  & \colhead{$\langle P_{gas,0} \rangle$}
  & \colhead{$M_s$} 
  & \colhead{$M_A$} 
  & \colhead{Resolution} 
}
\startdata
M1  & $0.1$ & $0.01$ & $\sim 7.6$ & $\sim 7.6$ & $512^3$ \\
M2  & $0.1$ & $0.1$  & $\sim 2.4$ & $\sim 7.7$ & $512^3$ \\
M3  & $0.1$ & $1.0$  & $\sim 0.8$ & $\sim 7.7$ & $512^3$ \\
M4  & $0.5$ & $0.01$ & $\sim 7.5$ & $\sim 1.5$ & $256^3$ \\
M5  & $0.5$ & $0.1$  & $\sim 2.3$ & $\sim 1.5$ & $256^3$ \\
M6  & $0.5$ & $1.0$  & $\sim 0.7$ & $\sim 1.4$ & $256^3$ \\
M7  & $1.0$ & $0.01$ & $\sim 7.4$ & $\sim 0.7$ & $512^3$ \\
M8  & $1.0$ & $0.1$  & $\sim 2.3$ & $\sim 0.7$ & $512^3$ \\
M9  & $1.0$ & $1.0$  & $\sim 0.7$ & $\sim 0.7$ & $512^3$ \\
M10 & $2.5$ & $0.01$ & $\sim 9.8$ & $\sim 0.4$ & $256^3$ \\
M11 & $2.5$ & $0.1$  & $\sim 3.3$ & $\sim 0.4$ & $256^3$ \\
M12 & $2.5$ & $1.0$  & $\sim 1.1$ & $\sim 0.4$ & $256^3$
\enddata
\end{deluxetable}

The initial magnetic field is of the form
$\mathbf{B}=\mathbf{B_{\mathrm{ext}}}+\mathbf{b}$, a uniform field
$\mathbf{B_{\mathrm{ext}}}$ plus a fluctuating part $\mathbf{b}$. 
Initially, $\mathbf{b}=0$ and $\mathbf{B_{\mathrm{ext}}}$ is aligned
in the $x-$direction. When the simulations reach a stationary state
the magnitude of the mean and the fluctuating parts are of the same
order, while the mean magnetic field remains aligned with the $x-$axis. 

\section{Velocity Anisotropy}
\label{sec:vel-anisotropy}

We have taken the results from the 3D-MHD simulations and produced 2D
maps of the mean velocity and centroids of velocity, integrating along 
each of the cardinal ($x~,y~,z$) directions.
We consider an isothermal, optically thin medium, with an emissivity
linearly proportional to the density (e.g. cold \ion{H}{1}). If we
integrate along the $x-$axis we can obtain a 2D map of the mean velocity
perpendicular to the $y-z$ plane:
\begin{equation}
  V_x(y,z)=\frac{1}{N_x}\int v_x(x,y,z)\,dx,
  \label{eq:mean-v}
\end{equation}
where $N_x$ is the number of cells used to discretize  $x$, and $v_x$
is the projection of the velocity field along $\mathbf{\hat{x}}$.
This mean LOS velocity traces the velocity structure in
a similar manner as the column density follows the density
structure (\citealt{2005ApJ...631..320E}, henceforth
\citetalias{2005ApJ...631..320E}).
However, it cannot be obtained directly from observations.

In real observations we are faced with a density weighted mean
(i.e. the velocity centroids).
For the centroid maps we used their conventional (normalized) form,
under the assumption of an optically thin medium whose emissivity is
proportional to the density (see \citetalias{2005ApJ...631..320E};
  \citealt{2007MNRAS.381.1733E}): 
\begin{equation}
C_x(y,z)=\frac{\int v_x(x,y,z)\,\rho(x,y,z)\,dx}{\int \rho(x,y,z)\,dx}.
 \label{eq:centroids}
\end{equation}
Analogous expressions to equations (\ref{eq:mean-v}) and
(\ref{eq:centroids}) can be used to obtain the mean velocity, or
velocity centroids with the LOS aligned with $\mathbf{\hat{y}}$, or
$\mathbf{\hat{z}}$. 

The two-point, second-order structure function of a quantity
$f(\mathbf{x})$ is defined as:
\begin{equation}
SF(\mathbf{r})=\left\langle \left[ f(\mathbf{x})-f(\mathbf{x}+\mathbf{r}) \right]^2 \right\rangle,
\label{eq:SF}
\end{equation}
where $\langle \hdots \rangle$ denotes an ensemble average over all
space ($\mathbf{x}$), and $\mathbf{r}$ is the separation or
``lag''. A closely related measure is the correlation function
${CF(\mathbf{r})=\left\langle
    f(\mathbf{x})f(\mathbf{x}+\mathbf{r})\right\rangle}$, which
differs from the $SF$ basically by a constant (see for instance
\citetalias{2005ApJ...631..320E}). 
For the sake of simplicity we will use the following notation:
$SF_{\mathrm{V,x}}(\mathbf{R})$, and 
$SF_{\mathrm{C,x}}(\mathbf{R})$ denote the structure function applied
to a map of mean velocity (eq. \ref{eq:mean-v}), and a map of
velocity centroids (eq. \ref{eq:centroids}), respectively. The lag
$\mathbf{R}$ 
is written in upper case letters to denote that is a two-dimensional
vector, while the $x$ in the sub-index indicates that the mean
velocity or centroids were obtained integrating along the $x-$axis [in
which case $\mathbf{R}=(y,\,z)$].
The power-spectrum is another recurrent tool in turbulence studies.
It is the Fourier transform of the correlation function, thus it
provides equivalent information. 

If the turbulence were isotropic these two point statistics would
(statistically) not depend on the direction of the lag, or on the
direction of the wave number for the power-spectrum. 
In fact it is customary to assume isotropy and average over all angles
to present just a radial dependence of $SF(r)$.
It is well known, however, that the magnetic fields breaks the
isotropy and the turbulence becomes anisotropic (see
\citealt{1984ApJ...285..109H} and references therein,
\citetalias{1995ApJ...438..763G}, \citealt{2005ThCFD..19..127C} for a
review).

\subsection{Additional Density Fluctuations}
\label{sec:fluct}

Spectroscopic observations are sensitive to density and velocity
fluctuations simultaneously. For the particular case of centroid maps,
it was clear for instance that the strong fluctuations in highly
supersonic turbulence (${M_s\gtrsim 2.5}$) affected severely our
ability to determine the spectral index of velocity
(\citealt{2003ApJ...592L..37L}; \citetalias{2005ApJ...631..320E};
\citealt{2007MNRAS.381.1733E}).

In order to further  study the impact of density fluctuations in the
anisotropy that can be observed we study four different maps obtained
for each of the simulations of Table \ref{tab:models}.
Firstly, a map of the mean velocity obtained as in equation
\ref{eq:mean-v}, which have information exclusively of velocity, but
can not be obtained from observations. Secondly, a map of velocity
centroids (eq. \ref{eq:centroids}), obtained with the density field
as obtained from the simulations.
Thirdly, another map of velocity centroids, but in this case
modulating the density by an $r^{-2}$ profile. The direction of the
gradient is $(\hat{x},\hat{y},\hat{z})$, so that it is oblique to
both the LOS and the mean B field. To avoid the singularity at $r=0$ we
have placed the origin half a pixel outside the computational domain,
and the resulting density field was rescaled to have a mean value of
$1$. 
This gradient introduce a large-scale density variation, 
which could be encountered in observations of the
ISM, but which is a not a product of turbulence (e.g. self-gravity).
And fourthly, another map of centroids was obtained with a density
field to which we have added white noise. We model the noise using
fractional Brownian motion (fBm) fractal structures, that can be 
characterized by a power spectrum index \citep{1998A&A...336..697S,
  2001A&A...366..636B}, which for white noise we have set to zero
(flat power spectrum). 
The resulting data cubes have a Gaussian probability distribution
function. Their dispersion have been scaled to the same value of
the original density for each of the models.
The noise is added to the density, and to preserve the density
positive defined we have set a floor value of $0.01$ (the mean density
is $1.0$).

\begin{figure}
\centering
\includegraphics[width=0.45\textwidth]{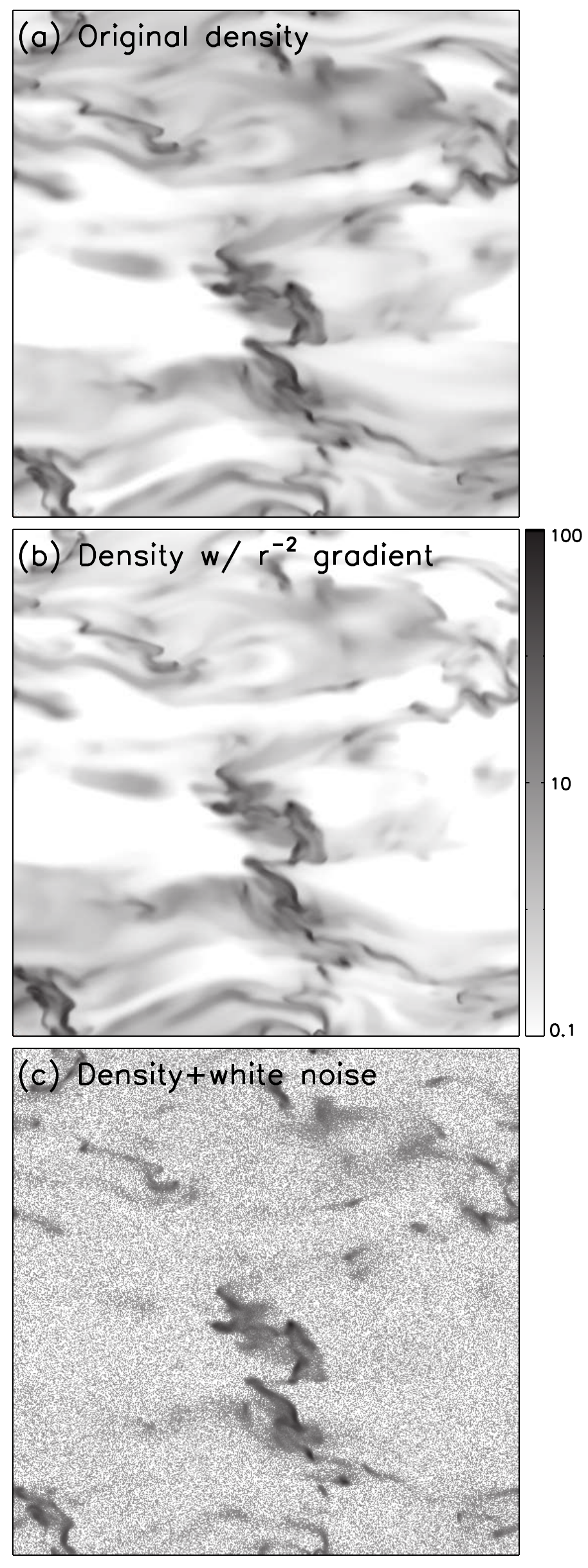}
\caption{Density cuts in the $XY-$midplane. (a) Original density, (b)
  density modulated by an $\propto r^{-2}$ gradient (decreasing from the left
  to the right, and from the bottom to the top), and (c) the density
  with additional white noise. All the plots are in the same
  (logarithmic) scale, as indicated by the bar on the right.
} 
\label{fig:maps}
\end{figure}

The resulting density fields are illustrated in Fig \ref{fig:maps},
where we present map of the $xy-$midplane cut of the density of one of
the models (M8, super-sonic and sub-Alfv\'enic,  see Table
\ref{tab:models}).  
The first panel (a) displays the original density, the
middle panel (b) shows the effect of the large scale gradient, and in
(c) the addition of white noise is evident.

In the following section we will study the correlation and structure
functions of the maps of mean velocity and of centroids, and how their
anisotropy depends on the parameters of the models.

\section{Results}
\label{sec:results}


Observations sample the entire LOS, and at a given velocity one has
contribution of material that could be anywhere along that LOS. In
some sense, one can say that observations average the information in
the position perpendicular to the plane of the sky. Thus, from an
observational point of view, it is more natural to study the
anisotropy in a global frame, namely, the anisotropy with respect to
the direction of the {\it mean} magnetic field (as opposed to the {\it local}
magnetic field, from which a \citetalias{1995ApJ...438..763G} scaling
is retrieved).

\begin{figure*}
\centering
\includegraphics[width=0.8\textwidth]{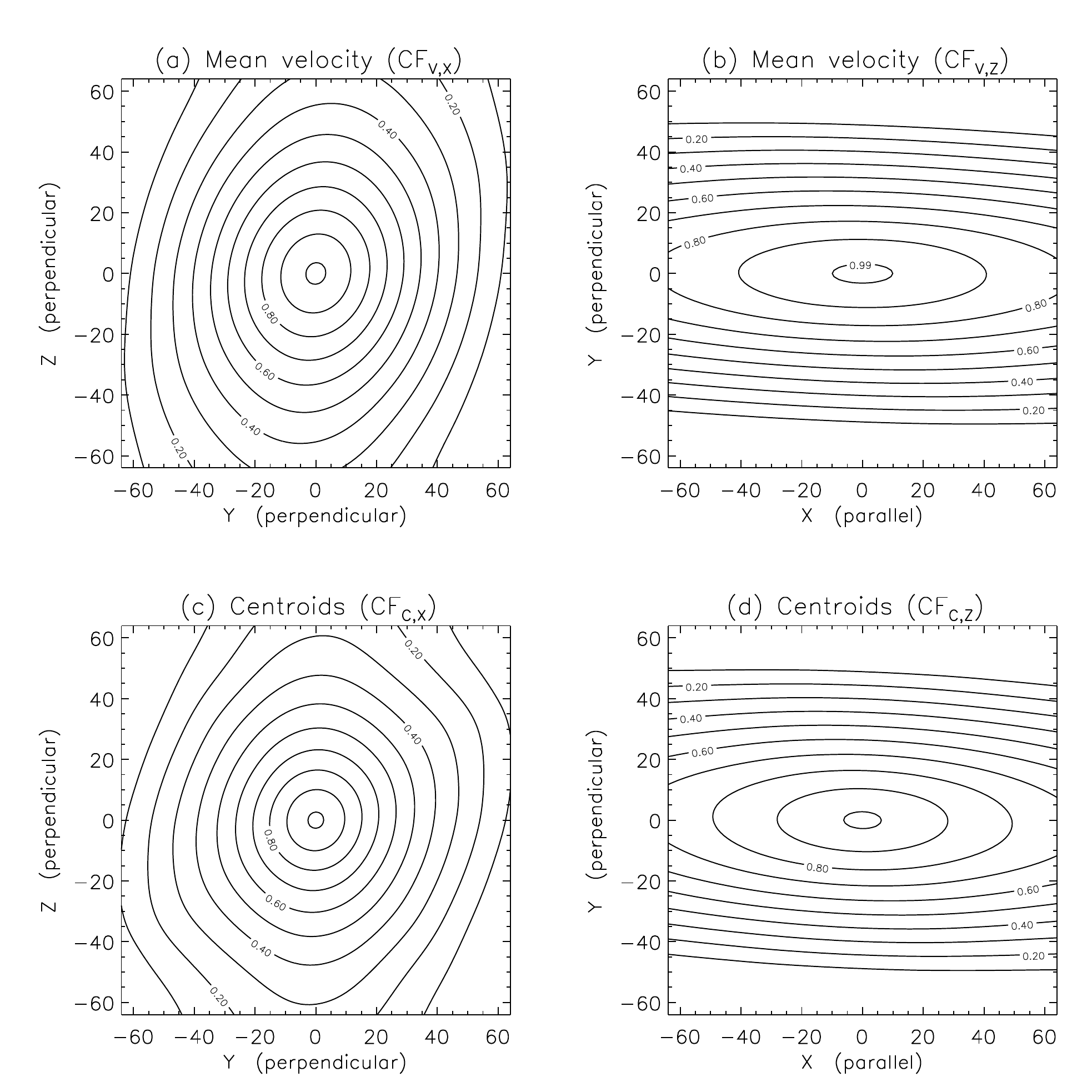}
\caption{Example of the iso-contours of the correlation function in
  one of the simulations (model M8:  super-sonic and sub-Alfv\'enic).
The first two panels [(a) and (b)] correspond to correlations in 
mean velocity maps, the last two [(c) and (d)] are correlations in
velocity centroid maps (with the original density, see \S\ref{sec:fluct}
).
In the plots on the left column [panels (a) and (c)] the LOS is
parallel to the mean B field. In the right column [panels (b) and (d)]
the LOS is perpendicular to the mean B field, which is in this case
aligned with the horizontal axis.
} 
\label{fig:contours}
\end{figure*}

We have shown (\citetalias{2002ASPC..276..182L};
\citealt{2003MNRAS.342..325E}, \citetalias{2005ApJ...631..320E}) that
in such a global frame indeed the statistics of velocity centroids reveal 
the direction of the mean magnetic field. 
In Figure \ref{fig:contours} we present contours of equal correlation
in one of the models (M8, the same used in Figure \ref{fig:maps}). 
The mean LOS contours are remarkably similar to those of velocity
centroids.
In the figure we only show the results for the centroids
with the original density because they are very similar to those
obtained if the $\propto r^{-2}$ density gradient, or white noise (see
\S\ref{sec:fluct}) are included.

In panels (a) and (c) in Figure \ref{fig:contours} the line of sight
is in the direction parallel to the mean magnetic field ($x$), thus the two
axes shown are perpendicular to it, and the contours are more or less
circular (isotropic). Panels (b) and (d) have been integrated in a direction
perpendicular to the mean plane ($z$) and the resulting correlations
show a clear anisotropy in the direction of the B field, the same
result is obtained if we integrate along the $y-$axis.
This result about the direction of the mean magnetic field (or more
accurately, its projection onto the plane of the sky) is quite
robust. 

\begin{figure*}
\centering
\includegraphics[width=1.\textwidth]{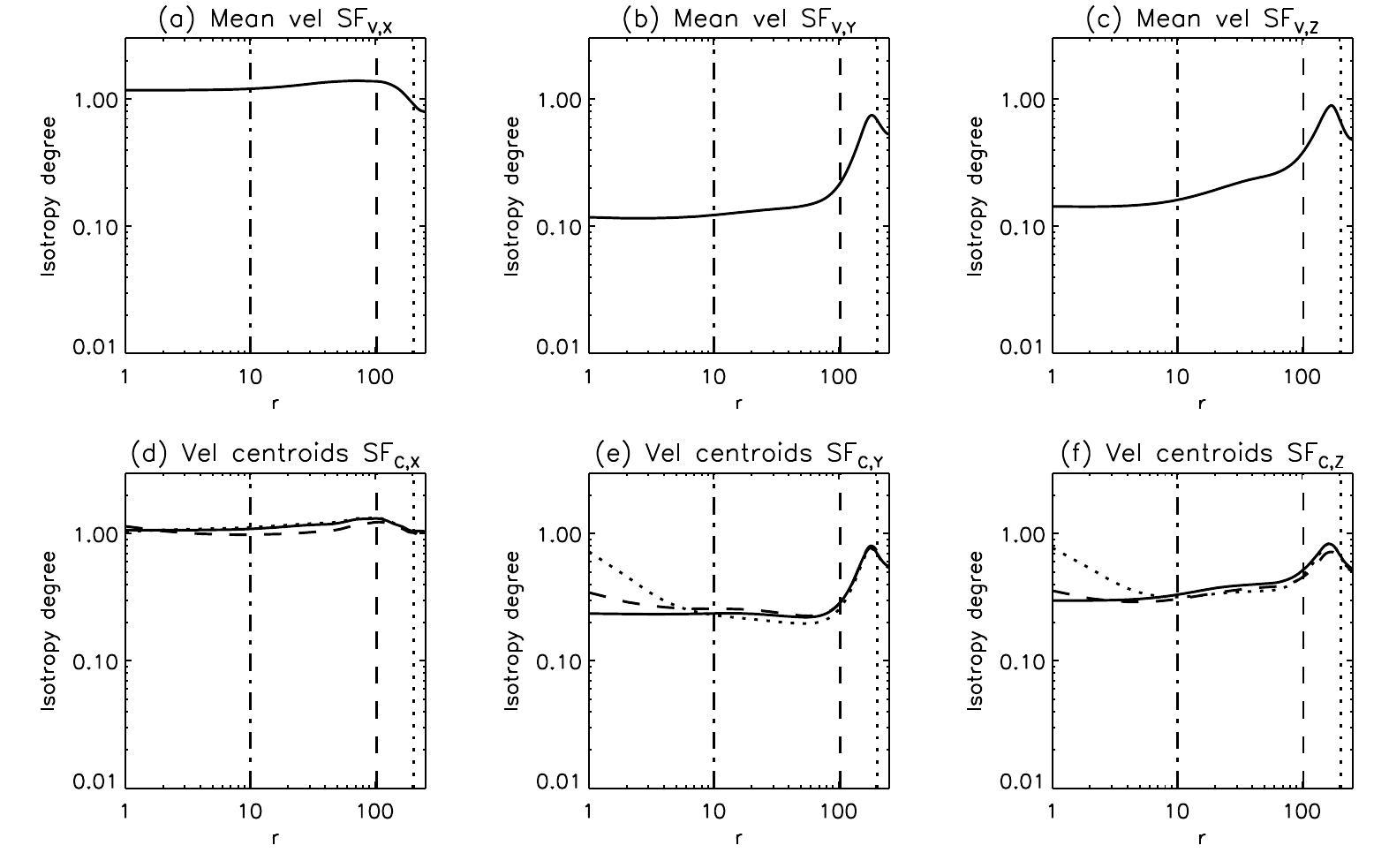}
\caption{Example of the degree of anisotropy of the structure functions
  in the same model shown in Figure \ref{fig:contours}, as observed
  from different directions (a value of 1 means isotropic).
The top row [panels (a)-(c)] are obtained with the mean density maps
while the bottom row [(d)-(f)] with maps of velocity centroids.
The different lines in panels (d-f) denote the density field used to
obtain the centroids, the {\it solid} line corresponds to the original
density, the {\it dashed} line to the $\propto r^{-2}$ gradient, and
the {\it dotted} line to the addition of white noise.  The LOS is
aligned with the $x-$axis (parallel to the B field) in the left column
[(a) and (d)], with the $y-$ axis in the middle column [(b) and (e)],
and with the $z-$axis in the right column [(c) and (f)]. 
}
\label{fig:ratios}
\end{figure*}

To address the dependence of the anisotropy with scale, we start by
defining a simple measure of the degree of anisotropy: the ratio of
the structure function in the two directions that 
are perpendicular to the LOS (e.g. $SF_{C,z}(x,0)/SF_{C,z}(0,y)$),
which we have computed for all the models.
In Figure \ref{fig:ratios} we present an example 
obtained from the same model of the previous two Figures. In the top
three panels (a-c) we show the results of the mean LOS velocity, and
in the bottom panels (d-f) the results with the different velocity
centroids.

If this ratio is one the structure function is isotropic, as in
the case of panels (a) and (d), which correspond to the structure
functions when the mean magnetic field and the LOS are aligned. The
rest of the panels are clearly anisotropic, with a degree of
anisotropy, whose exact value depended on model. 
In Figure \ref{fig:ratios} we indicated with vertical lines the scale
length of injection (turbulence forcing) with a dotted line. At such
large scales, and down to about  $1/5$ of the box size (marked with a
vertical dashed line) the effect of the forcing is evident. 
For the mean LOS velocity, and the centroids with the original density
field, we found  that the anisotropy was virtually scale independent
from the small scales up to separations on the order $\sim 1/5$ of the
computational box (half the size of the injection scale). 
For centroids with additional density fluctuations (dashed and dotted
lines) we see that the anisotropy shows a flat (scale independent)
behavior for separations $\gtrsim 10$ and up to  $\sim 1/5$ of the
computational box. In other words, the additional density fluctuations
interfere with the measured anisotropy degree at the smallest scales,
making the statistics more isotropic.

\begin{figure*}
\centering
\includegraphics[width=1.\textwidth]{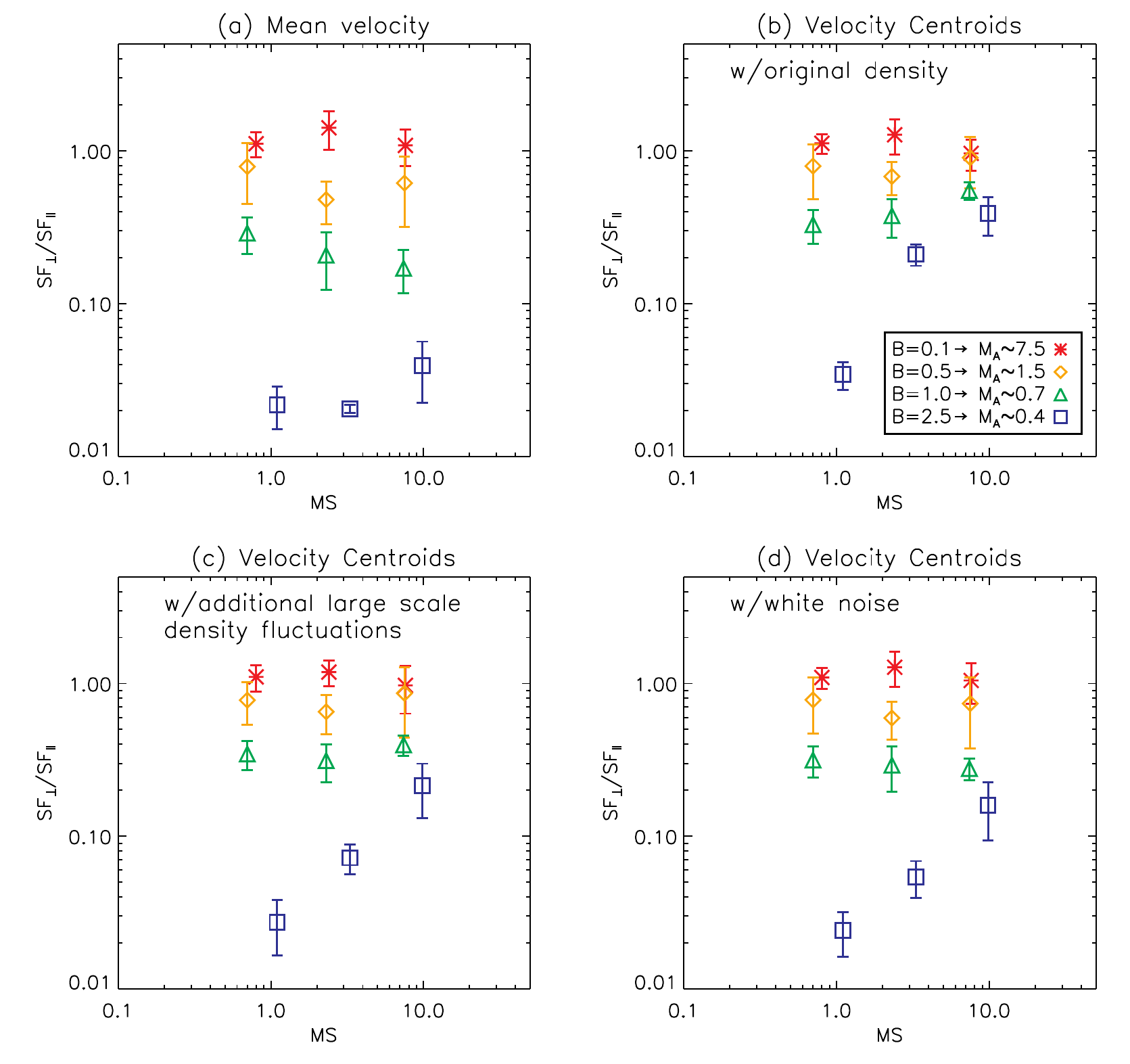}
\caption{
  Degree of anisotropy in all the models averaged over scales
  from 10 grid points to $1/5$ of the computational box. 
  The horizontal axis corresponds to the sonic Mach Number, the
  Alfv\'enic Mach Number is indicated by the various symbols (and
  colors in the online version) as shown in the label. 
  In the left panel (a) we plotted the anisotropy in velocity
  centroids, in the right panel (b) the anisotropy in the average LOS
  velocity. In both figures the results are obtaining averaging
  the two cases where the LOS is perpendicular to the mean
  field. The error bars show the maximum variation of the averaging
  procedure (including variation across scales).}
\label{fig:ratios2}
\end{figure*}

More interesting than confirming the scale independence on the
structure functions in the inertial range of the turbulent cascade
(there is already evidence of this in
\citetalias{2002ApJ...564..291C}), it is to study how the degree of
anisotropy  depends on the turbulence parameters (sonic and Alfv\'enic
Mach numbers at the injection scale). 
To do this we have calculated the degree of anisotropy on all the
models, as exemplified for model M8 in Fig. \ref{fig:ratios} and
computed the average value in scales between 10 grid points  and
$L/5$. 
Below 10 grid points scale, the density from the MHD simulations is
severely affected by numerical diffusion, and the effect of the large
scale density gradient and noise are more pronounced.
The results are condensed in Figure \ref{fig:ratios2}, where we have
plotted the average degree of isotropy as a function of the sonic Mach
number and of the Alfv\'enic Mach number. 
The Mach numbers are indicated by the different symbols (and colors in
the online version) as displayed in the legend. The error bars show
the variations while obtaining the average value.  They are the product
of differences in the two possible lines of sight ($y$- and $z$-axis) and 
slight scale variations. 

One can see from Figure  \ref{fig:ratios2} that the centroids 
maps without noise or gradients have the smallest error bars, thus the least
dependence on scale. The rest of the centroids and the mean LOS
velocity have slightly larger error bars, and therefore some (small)
scale dependence.
It is also quite noticeable that, the results for the different
centroids (with the original density or modified data) are very
similar. This is not too surprising, because we have
restricted ourselves to scales that were seen as less effected by
either the $\propto r^{-2}$ gradient or the noise.
However, to apply this technique to real data it would be advisable to
calculate the anisotropy degree on several scales and search for a
scale independent range, in the same manner one restricts the inertial
range looking for power-laws when analyzing power spectra.

It is clear also, that the degree of anisotropy depends mostly on the
Alfv\'enic Mach number: for increasing values of the magnetic field
the level of anisotropy increases as well (smaller
$SF_{\perp}/SF_{\parallel}$ ).
For the velocity centroids we have the same general trend, but not as
pronounced. For instance, for $M_A\gtrsim1.4$ the degree of
anisotropy is barely distinguishable from isotropic, while in the mean
LOS velocity map this was only the case for a very weak B field ($M_A\sim7.5$).
Centroids also show a weak dependence on $M_s$, but only for
moderate to low magnetizations ($M_A>0.4$).
For strong magnetic fields ( $M_A\sim0.4$) we observe
some dependence on  the sonic Mach number.
Since this was not noticeable in Fig.\ref{fig:ratios2}(a), and given
the similarity of Figs. \ref{fig:ratios2}(b-d), one can attribute such
dependence to the original density field (i.e. arising from shocks in
supersonic turbulence).
This strong influence of the sonic Mach number is seen as a positive
slope for the most magnetized simulations.

The results of a clean dependence on the Alfv\'enic Mach number and a
weak dependence on $M_s$ (for small to moderate magnetic field
strengths), are encouraging. While the sonic Mach number can 
be obtained by a variety 
of techniques \citep{1997ApJ...474..730P, 2003A&A...398..845P,
  2007ApJ...658..423K, 2008ApJ...688L..79F, 2009ApJ...693..250B}, the
Alfv\'enic Mach number has remained elusive. We are confident that
anisotropy studies, along with the Chandrasekhar-Fermi technique
\citep[see for instance][]{2008ApJ...679..537F} are starting to change
this situation.

All in all, the observable turbulent fluctuations, as represented by
velocity centroids, are sensitive to the fluid magnetization given by
the Alfv\'enic Mach number $M_A$. The dependence of the anisotropy on
the sonic Mach number $M_s$ is, however,  not always negligible. The
latter number can be obtained using other statistics studied in the
literature \citep[see][]{2007ApJ...658..423K, 2009ApJ...693..250B,
  2010ApJ...708.1204B, 2010ApJ...710..125E}

\section{Discussion and Summary}
\label{sec:summary}

Advances in understanding of the nature of magnetized turbulence drive
the development of the techniques to study turbulence  through
observations. For these studies different approaches can (and should)
be used. However, it is important to understand what is the maximal
information that one can get from observations. 

Recently observational studies of interstellar anisotropies have been
performed by \cite{2008ApJ...680..420H}. The resulting anisotropies
were broadly consistent with the expectations obtained in theoretical
and numerical studies (see \citetalias{2002ASPC..276..182L,
  2005ApJ...631..320E}), revealing the direction of the mean magnetic
field.
Both theoretical predictions and numerical calculations,
including those in the present paper indicate that the anisotropy
expected in the global frame of reference, which is sampled in the
observations, should have a range in which they do not depend on the
scale. In addition, we pay attention to the dependence of the velocity
anisotropy on the sonic Mach numbers.

In general, the measures we use to study the anisotropy may be
affected by other interfering factors. For instance, density effects
can affect velocity centroid measures. Attempts to mitigate the effect
fluctuating density field on velocity centroids was attempted in
\citet{2003ApJ...592L..37L} where new measures termed ``modified
centroids'' were introduced. However, further studies in
\citet{2005ApJ...631..320E, 2007MNRAS.381.1733E} revealed that the
``modified centroids'' improve the representation of turbulent
velocity only at moderate $M_s$. We have also used MVCs for the models
presented in this paper and found that the results were actually worse
(noisier) than with ordinary centroids.
Thus in the present study of anisotropy we present only ordinary
centroids. 

To study the robustness of the method we have added fluctuations to
the density field and reanalyze the anisotropy of centroids obtained
with these new data. We find that the range of scales in 
which the anisotropy degree is scale independent is limited to
separations smaller than the injection scale (in fact smaller than
half of the injection scale), but at the same time large enough to
avoid the effects of the additional fluctuations. In our models we
have found that the effect of an $\propto r^{-1}$ gradient and white
noise can be avoided by setting the smallest scale used to measure the
anisotropy to $10$ cells. With real observations one has to measure
the anisotropy and determine the appropriate range. 

\citet{2008ApJ...680..420H}, use not centroids, but a Principal
Component Analysis (PCA) to study turbulence through
observations. PCA analyzes the data in position-position-velocity
(PPV) space which could potentially have some useful information that
is lost in the averaging procedure in velocity centroids. However, PCA
relies in calibration from numerical models, while structure functions
and centroids can be described analytically and predictions of their
results follow directly from theory.

Our simulations show a dependence on the sonic Mach number which was
not reported by \cite{2008ApJ...680..420H}.  We suspect the reason is
that dependence on $M_s$ is less prominent compared
to the one on $M_A$. However, if one wants a more precise analysis it
should be taken into account.
In spite of this, we should stress the pioneering
significance of the observational studies of
\cite{2008ApJ...680..420H} which moved this technique of magnetic field
study from theoretical and numerical domain (see
\citetalias{2002ASPC..276..182L, 2005ApJ...631..320E}) to a
domain of practical application. We believe that the technique has
great future and view this paper as a contribution to its improvement.

We have taken a set of simulations of fully-developed, driven MHD
turbulence with different combinations of sonic and Alfv\'enic Mach
numbers to study the velocity anisotropy available from observations.

Our results can be summarized as follows:
\begin{itemize}

\item Synthetic maps obtained through compressible
  MHD simulations reveal a clear anisotropy of the velocity field, in
  alignment with the direction of the mean magnetic field. This is true
  for Alfv\'enic Mach numbers less than $M_A\simeq  1.5$.

\item The anisotropy, measured at scales in the inertial range of the
  turbulent cascade is {\it scale independent} in the global frame
  of reference, as opposed to the scale dependent anisotropy with is
 obtained with respect to the the direction of the local magnetic
 field.

\item The degree of anisotropy is dominated by the Alfv\'enic Mach
  number. However, a dependence on the sonic Mach number gets
  prominent for highly magnetized gas. The anisotropy in velocity
  centroids differs from that of the mean velocity. As many 
  other statistics are sensitive to the sonic Mach number this allows
  to make a correction for  $M_s$ to retrieve $M_A$.

\item To show that the method is robust against large-scale density
  fluctuations and noise, we have obtained centroids with modified
  density fields, to which we have added an $\propto r^{-2}$ gradient,
  or white noise. We see that these have an effect that is mostly seen
  at the smallest scales. If the degree of anisotropy is measured
  avoiding such scales the results remained basically unchanged.

\end{itemize}

\acknowledgments

AE acknowledges support from grants CONACyT 61547, 101356, and
101975. AL acknowledges the NSF grant AST 0808118 and the support of
the Center of Magnetic Self-Organization (CMSO). 

\bibliography{Master}

\end{document}